\documentclass[11pt]{article}

\usepackage{geometry}                
\usepackage{adjustbox}
\usepackage{booktabs}
\usepackage[table]{xcolor}
\usepackage{siunitx}
\usepackage{amsmath}
\usepackage{array}
\usepackage{graphicx}
\usepackage{amssymb}
\usepackage{epstopdf}
\usepackage{float}
\usepackage[english]{babel}
\usepackage[latin1]{inputenc}
\usepackage{filecontents}
\usepackage{caption}
\newtheorem{theorem}{Theorem}

\usepackage{authblk}

\begin{document}

\title{The role of disease cycles in the endemicity of infectious diseases}

\date{}

\author[1]{Esteban Vargas}
\author[1]{Camilo Sanabria}

\affil[1]{Department of Mathematics, Universidad de los Andes, Bogota, Colombia}

\maketitle

\begin{abstract}
Vector-borne diseases with reservoir cycles are complex to understand because new infections come from contacts of the vector with humans and different reservoirs. In this scenario, the basic reproductive number $\mathcal{R}^h_0$ of the system where the reservoirs are not included could turn out to be less than one, yet, an endemic equilibrium be observed. Indeed, when the reservoirs are taken back into account, the basic reproductive number $\mathcal{R}_0^r$, of only vectors and reservoirs, explains the endemic state. Furthermore, reservoirs cycles with a small basic reproductive number could contribute to reach an endemic state in the human cycle. Therefore, when controlling for the spread of a disease, it could not be enough to focus on specific reservoir cycles or only on the vector. In this work, we created a simple epidemiological model with a network of reservoirs where $\mathcal{R}_0^r$ is a bifurcation parameter of the system, explaining disease endemicity in the absence of a strong reservoir cycle. This simple model may help to explain transmission dynamics of diseases such as Chagas, Leishmaniasis and Dengue. 
\end{abstract}


\section{Introduction}

Some tropical diseases are amplified by one or several reservoirs. This is the case in diseases such as Chagas disease and Leishmaniasis. Indeed, Chagas disease has a domiciliary cycle, where domestic animals act as reservoirs, and a sylvatic cycle, where mammals like rodents are reservoirs \cite{paniker2007textbook}. Regarding Leishmaniasis, the main reservoirs of the disease in countries of South America are dogs, but other mammals could also act as reservoirs. In this paper, we are interested in diseases that have a network of reservoirs. We are also interested in representing those diseases in a simple mathematical model where we can measure the amplification effects of the reservoirs through the basic reproductive number.

In mathematical models of infectious diseases based on ordinary differential equations, the basic reproductive number of the disease is frequently obtained using the method of the Next Generation Matrix (NGM) presented in \cite{van2002reproduction}. Different interpretations of the NGM can lead to different basic reproductive numbers. In Section \ref{sappendix} we present the construction of the NGM that it is used in this work. 

\begin{figure}
\includegraphics[scale=0.5]{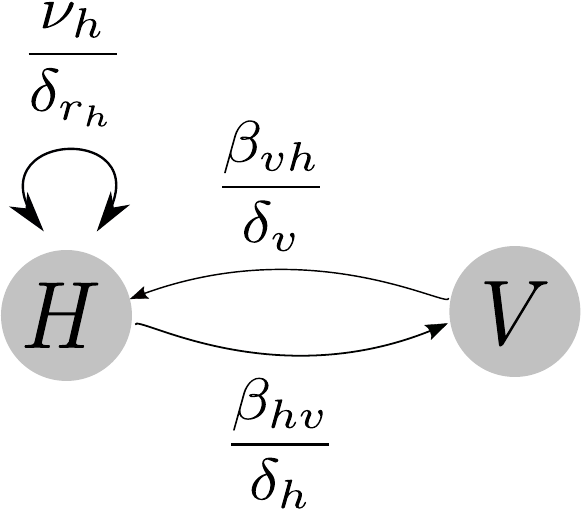}
\caption{The node $V$ represents the infectious vector and the node $H$ represents the infectious humans. The expressions next to the arrows represent the number of infections in the ending node caused by and individual in the initial node during its generation.}
\label{feje1}
\end{figure}

As an example, we consider the system represented in Figure \ref{feje1} and equations (\ref{eje1}). This system represents the transmission of a disease between vectors $V$ and humans $H$ with transmission rates $\beta_{vh}$, $\beta_{hv}$ and mortality rates $\delta_v$ and $\delta_h$. The disease can also be transmitted among humans with transmission rate $\nu_h$. This model assumes that both populations are constant, so the model is determined by the equations of the infectious populations in (\ref{eje1}).  

\begin{equation}
\begin{cases}
\dot{I_h} = \beta_{vh} I_v \frac{S_h}{N_h}+ \nu_h I_h \frac{S_h}{N_h}  - \delta_h I_h\\
\dot{I_v} = \beta_{hv} I_h\frac{S_v}{N_v} - \delta_v I_v\\
\end{cases}
\label{eje1}
\end{equation}

The basic reproductive number that is obtained using the NGM depends on the interpretation of which infections are considered as new. In the system presented above we could defined human and vector infections as new infections, or only human or vector  infections as new. From these three interpretation we get three basic reproductive numbers (see Subsection \ref{ssngm}). From the Theorem  \ref{umbral} that is proven in \cite{van2002reproduction}, these three numbers are greater than one (in this case the disease free equilibrium is locally asymptotically stable), or the three numbers are less than one (in this case the disease free equilibrium is unstable). In consequence, to check the stability of a possible endemic state of a system we could take an appropriate interpretation of NGM guided by the simplicity of the calculations.   

In Section \ref{smodel} we propose an epidemiological model of a vector-borne disease that has a network of reservoirs that infect one another. In Section \ref{sresults} we show the basic reproductive number of the simplified system (omitting infections between different reservoirs) in terms of the basic reproductive number of the human cycle and the reservoirs cycles.  We also present an application to Chagas disease based on data in Colombia taken from \cite{cordovez2014environmental}. It is shown that the disease is getting extinct as long as synergistic control is made in the number of vectors and reservoirs. In Section \ref{sdiscusion} we present the discussion and conclusions of the results presented in the Section \ref{sresults}. In Section \ref{sappendix} we present the method of the NGM and the mathematical justification of results in Section \ref{sresults}.


\section{The model} \label{smodel}

\begin{figure}[H]
\includegraphics[scale = 0.5]{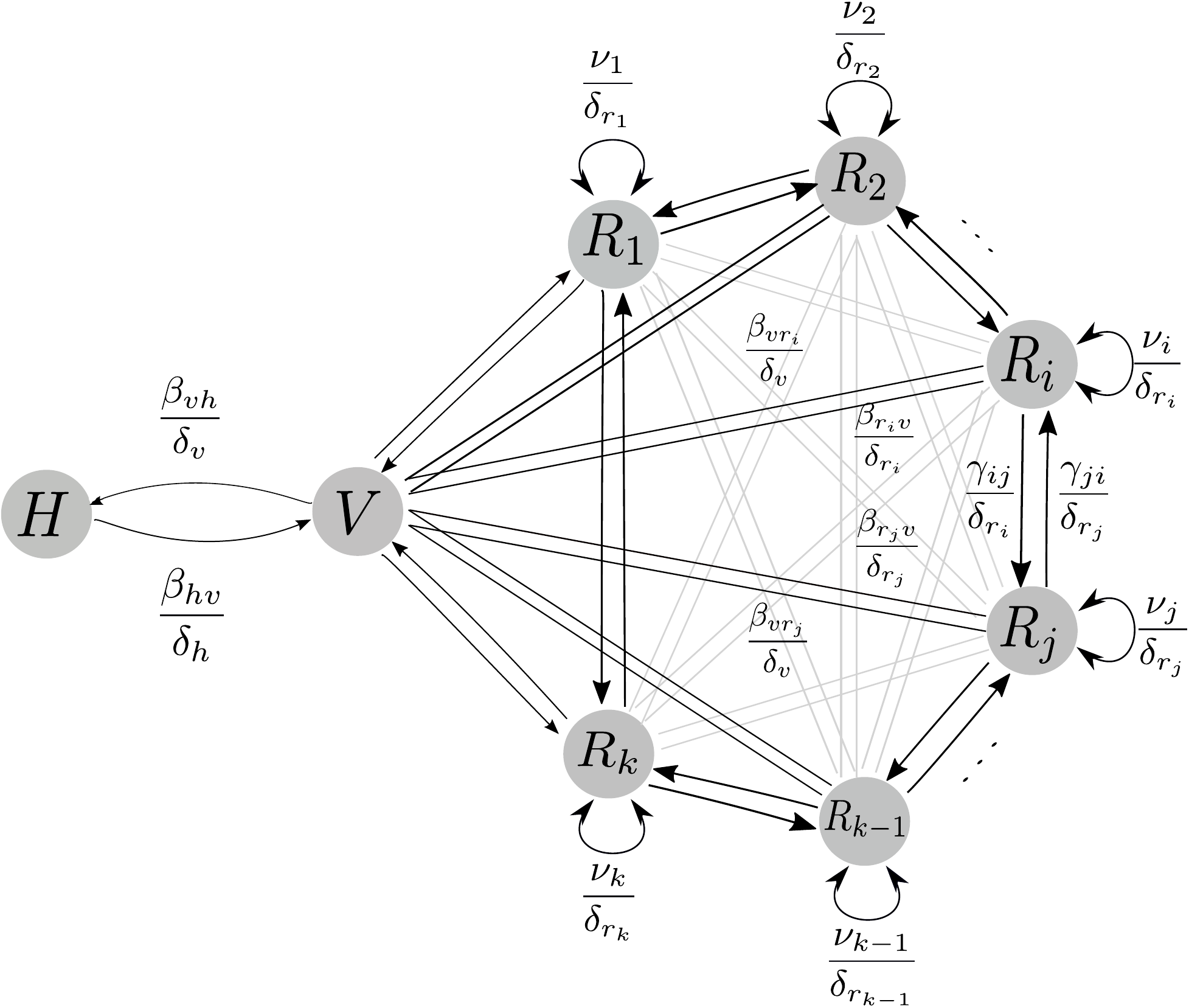}
\caption{The node $V$ represents the infectious vector, the node $H$ represents the infectious humans and the nodes $R_i, i=1,\ldots,k $ represent the infectious reservoirs. The weights next to the arrows represent the number of infections in the ending node caused by and individual in the initial node during its generation. }
\label{complete}
\end{figure}

We propose a mathematical model of a vector-borne disease that has a network of $k$ reservoirs. The state variables of the system are the Human population ($H$), the vector ($V$) and the reservoirs ($R_i, i= 1, \ldots,k $). We suppose that all the populations are constant ($N_h$ humans, $N_v$ vectors and $N_{r_i}$ reservoirs of the species $R_i$, $ i= 1, \ldots,k $).
We assume that in each reservoir species there could be self infection. Besides, the reservoirs can infect one another but there is no infection between reservoirs and human as the lines in Figure \ref{complete} shows. The parameters of the model are presented in Table \ref{param1} and the system of differential equations for the infectious populations of humans, reservoirs and vectors ($I_h,I_{r_i}, I_v$  respectively) that describes the model is given in (\ref{ecompleto}).

\begin{table}
\caption{Parameters of the model described by equations (\ref{ecompleto}).}
\label{param1}
\begin{tabular}{llll}
\hline\noalign{\smallskip}
Parameter & meaning & Units \\ 
\noalign{\smallskip}\hline\noalign{\smallskip}
$\beta_{vh}$ & Number of human infections caused by& $[H]/([time]*[V])$ \\
& one infectious vector per unir of time & & \\
$\beta_{hv}$ & Number of vector infections caused by & $[V]/([time]*[H])$ \\
& one infectious human per unir of time &  & \\
$\beta_{vr_i}$ & Number of infections of reservoir $R_i$ caused by  & $[R_i]/([time]*[V])$ \\
& one infectious vector per unir of time & &\\
$\beta_{r_iv}$ & Number of vector infections caused by  & $[V]/([time]*[R_i])$\\
& one infectious reservoir $R_i$ per unir of time & \\
$\gamma_{ij}$ & Number of infections of reservoir $R_j$ caused by  & $[R_j]/([R_i]*[time])$ \\
& one infectious reservoir $R_i$ per unir of time & \\
$\delta_h$ & Mortality rate of humans & $1/[time]$ \\
$\delta_v$ & Mortality rate of vectors & $1/[time]$ \\
$\delta_{r_i}$ & Mortality rate of reservoirs $R_i$ & $1/[time]$ \\

\noalign{\smallskip}\hline
\end{tabular}
\end{table}

\begin{equation}
\begin{cases}
\dot{I_h} = \beta_{vh} I_v \frac{S_h}{N_h} - \delta_h I_h\\
\dot{I_{r_i}} = \beta_{vr_i} I_v \frac{S_{r_i}}{N_{r_i}} + \nu_i I_{r_i} \frac{S_{r_i}}{N_{r_i}} +\sum_{j\neq i}\gamma_{ji} I_{r_j}\frac{S_{r_i}}{N_{r_i}} - \delta_{r_i} I_{r_i}, i = 1, \ldots, k\\
\dot{I_v} =    \beta_{hv} I_h\frac{S_v}{N_v} +\sum_{i=1}^k \beta_{r_iv} I_{r_i}\frac{S_v}{N_v} - \delta_v I_v\\
\end{cases}
\label{ecompleto}
\end{equation}

\section{Results} \label{sresults}

\subsection{Basic reproductive number of simplified model} \label{ssr0}

For the graph in Figure \ref{complete} we define the weight of each edge as the expression next to it. For example, the weight of the edge from $H$ to $V$ is $\beta_{hv}/\delta_h$. The meaning of the weight of the edge from node $x$ to node $y$ is the number of infections in node $y$ that one individual of node $x$ can cause during its generation. For a cycle of the graph, we say that its weight is the geometric mean of the weights of its edges. For example, the two nodes cycle formed by $H$ and $V$ has a weight $\sqrt{\beta_{vh}/\delta_v\beta_{hv}/\delta_h}$. We denote the weight of the cycle with nodes $x$ and $y$ by $\mu_{xy}$. The following result due to Friedland \cite[Theorem 2]{friedland1986limit} gives us upper and lower bounds of the basic reproductive number in terms of the weights of the cycles of the graph.

\begin{theorem}
Let $A$ be a matrix with nonnegative entries and $$\Omega := \{ \sigma: \{ i_1 \ldots i_k\} \rightarrow \{i_1, \ldots, i_k\}: \sigma \textmd{ is a cycle}, \{i_1, \ldots, i_k\} \subseteq \{1, \ldots, n\}\}.$$ For $\sigma \in \Omega$, we define  $\mu_\sigma := (\prod _ {j=i}^k a_{i_j,\sigma(i_j)})^{1/k}$. If $\mu_*(A) := max_{\sigma \in \Sigma} \{\mu_\sigma\}$ and  $S(A)  := (Sign(a_{i,j}))$, then
$$lim_{r \rightarrow \infty} \mu(A) := \rho(A^{[r]})^{1/r} = max_{\sigma \in \Omega_n} \mu_\sigma(A)$$ and $$\mu(A) \leq \rho(A) \leq \rho(S(A)) \mu(A).$$ 
\label{teoumbral}
\end{theorem}

If $G$ is the NGM of an epidemiological model, then $S(G)$ determines the pairs of species where there is infection. Moreover, if $\mu_*$ is the heaviest cycle, we get that:

\begin{equation}
\mu_*(G) \leq \mathcal{R}_0 \leq \rho(S(G)) \mu_*(G) 
\label{cotag}
\end{equation}

In particular, this shows that a cycle with node $x$ and $y$ has basic reproductive number greater than one, i.e.,  $\mu_{xy} >1$, then the basic reproductive number of the whole system is also greater than one.

Constructing the NGM of the model presented in Section \ref{smodel} as it is explained in Subsection \ref{ssngm} of Appendix, we obtain that if we consider the infection of all species as new, the matrices $F$ and $V$ that define the NGM are:

$$F=
\begin{pmatrix}
0 & \beta_{vh} & 0 & 0 & \ldots & 0\\
\beta_{hv} & \nu_v &  \beta_{r_1v}  & \beta_{r_2v} & \ldots & \beta_{r_kv} \\
0 & \beta_{vr_1} & \nu_{r_1} & \gamma_{21} & \ldots & \gamma_{k1} \\
0 & \beta_{vr_2} & \gamma_{12} & \nu_{r_2} & \ldots & \gamma_{k2} \\
\vdots & \ddots & \ddots & \ddots & \ldots &\gamma_{kk-1}\\
0 & \beta_{vr_k} & \gamma_{1k} & \gamma_{2k} & \ldots & \nu_{r_k} \\

\end{pmatrix}, V=
\begin{pmatrix}
\delta_h & 0& 0 & 0 & \ldots & 0\\
0 &  \delta_v &  0 & 0 & 0 & 0 \\
0 & 0 & \delta_{r_1} & 0 &  0 & 0 \\
0 & 0 & 0 & \delta_{r_2} & \ldots & 0 \\
\vdots & \ddots & \ddots & \ddots & \ldots &0\\
0 & 0 & 0 & 0 & \ldots & \delta_{r_k} \\

\end{pmatrix}$$

In consequence, the NGM of the system is:

\begin{equation}
G = FV^{-1}=
\begin{pmatrix}
0 & \frac{\beta_{vh}}{\delta_v} & 0 & 0 & \ldots & 0\\
\frac{\beta_{hv}}{\delta_h} &  \frac{\nu_v}{\delta_v} &  \frac{\beta_{r_1v}}{\delta_{r_1}} & \frac{\beta_{r_2v}}{\delta_{r_2}} & \ldots & \frac{\beta_{r_kv}}{\delta_{r_k}} \\
0 & \frac{\beta_{vr_1}}{\delta_{v}} & \frac{\nu_{r_1}}{\delta_{r_1}} & \frac{\gamma_{21}}{\delta_{r_2}} & \ldots & \frac{\gamma_{k1}}{\delta_{r_k}} \\
0 & \frac{\beta_{vr_2}}{\delta_{v}} &  \frac{\gamma_{12}}{\delta_{r_1}} & \frac{\nu_{r_2}}{\delta_{r_2}} & \ldots & \frac{\gamma_{k2}}{\delta_{r_k}} \\
\vdots & \ddots & \ddots & \ddots & \ldots &\frac{\gamma_{kk-1}}{\delta_{r_k}}\\
0 & \frac{\beta_{vr_k}}{\delta_{v}} & \frac{\gamma_{1k}}{\delta_{r_1}}  & \frac{\gamma_{2k}}{\delta_{r_2}}  & \ldots & \frac{\nu_{r_k}}{\delta_{r_k}} \\
\end{pmatrix}
\label{gsimple2}
\end{equation}

Let us consider the system presented in the previous section with $\gamma_{ij}=0$ for all $i,j \in \{1,\ldots, k\}, i\neq j$. In this case, the spectral radius of the matrix $G$ in (\ref{gsimple2}) is the greatest root of the equation in (\ref{pcsimple2}) for $\lambda$.

\begin{equation} \label{pcsimple2}
\lambda( (\nu_v / \delta_v - \lambda) - \sum_{i=1}^n \frac{\mu_{vr_i}^2}{\nu_{r_i}/\delta_{r_i} - \lambda } )= -\mu_{vh}^2
\end{equation}

In general, equation (\ref{pcsimple2}) is not easy to solve. However, if we omit self infection in all reservoirs, i.e., $\nu_{v}=0, \nu_{h}=0, \nu_{r_i}=0$ for $i=1,\ldots,k$, we get that the greatest solution of (\ref{pcsimple2}) is given by (\ref{r0msimple}).

\begin{equation} \label{r0msimple}
 \mathcal{R}_0=\rho(FV^{-1}) =  \sqrt{\frac{\beta_{vh}}{\delta_v} \frac{\beta_{hv}}{\delta_h} + \sum_{i=1}^k \frac{\beta_{vr_i}}{\delta_v} \frac{\beta_{r_iv}}{\delta_{r_i}} } = \sqrt{\mu_{vh}^2 + \mu_{vr_1}^2 + \ldots + \mu_{vr_k}^2  }
\end{equation}

In this scenario, the set of cycles $\Omega$  would only have two nodes cycles. Moreover, If $\mu_* := max_{\sigma \in \Omega} \mu_{\sigma} = max \{\mu_{vh}, \mu_{vr_1},\ldots,\mu_{vr_k}\} $, the inequalities in (\ref{cotag}) would give us the obvious bounds in (\ref{cotasimple1}).

\begin{equation}\label{cotasimple1}
\mu_*^2 \leq \mu_{vh}^2 + \mu_{vr_1}^2 + \ldots + \mu_{vr_k}^2 \leq (k+1) \mu_*^2
\end{equation}

If we take into account the self infection in all reservoirs, the inequalities in (\ref{cotag}) would turn into the inequalities in \ref{umbralsimple2} if $$\mu_* = max \{\mu_{vh}, \mu_{vr_1},\ldots,\mu_{vr_k}, \nu_v/\delta_v, \nu_{r_1}/\delta_{r_1}, \ldots, \nu_{r_k}/\delta_{r_k}\}.$$ 

\begin{equation}
 \mu_*  \leq \mathcal{R}_0 \leq \rho(S)\mu_* = (\sqrt{k+1}+1) \mu_*
 \label{umbralsimple2}
\end{equation}

If $\mu_*>1$, the disease free equilibrium would be unstable (see Theorem \ref{umbral} in Appendix). If $(\sqrt{k+1}+1) \mu_*<1$, the disease free equilibrium would be locally asymptotically stable. Nonetheless, the inequalities in  (\ref{umbralsimple2}) does not let us determine whether $\mathcal{R}_0<1$ or $\mathcal{R}_0>1$ when $(\sqrt{k+1}+1) \mu_*>1$ and $\mu_*<1$. To solve this problem we interpret the NGM to obtain matrices $F$ and $V$ that ease the computation of the spectral radius of $FV^{-1}$.

\begin{figure}
\includegraphics[scale=0.5]{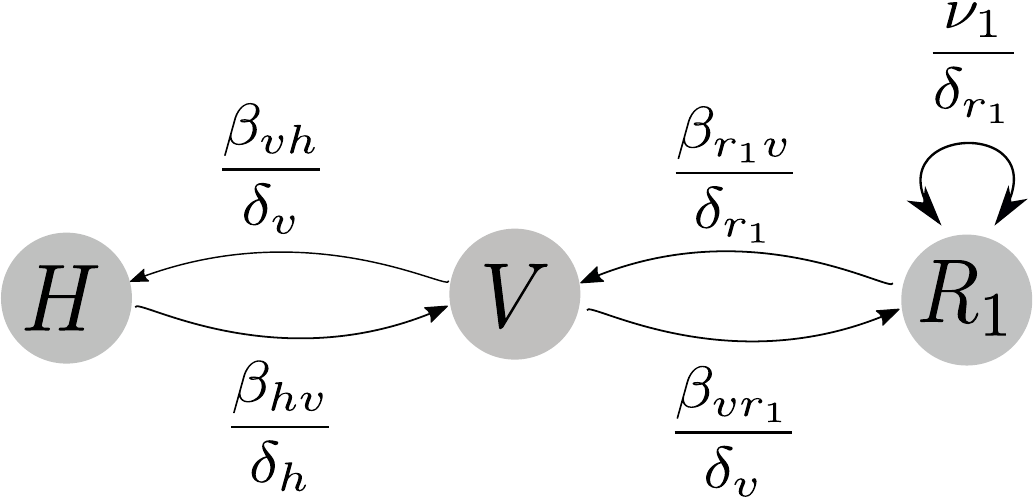}
\caption{Simplification of the model presented in Section \ref{smodel} considering only one reservoir.}
\label{HVr1}
\end{figure}

For simplicity of the explanation, let us consider the model presented in Section \ref{smodel} with only one reservoir, as the graph in  Figure \ref{HVr1} represents.

We define the threshold values in Table \ref{tvalues1} from their respective interpretation of the next generation matrix.

\begin{table}
\caption{Umbral values defined from the systems formed by the state variables of the second column and the interpretations of the next generation matrix given by the third column.}
\label{tvalues1}
\begin{tabular}{llll}
\hline\noalign{\smallskip}
Value & System form by & New infections \\ 
\noalign{\smallskip}\hline\noalign{\smallskip}
$\mathcal{R}_0$ & $H,V,R$ & $H$ \\
$\mathcal{R}_0^r$ & $V, R$ & $V$ \\
$\mathcal{R}_0^h$ & $H,V $& $H$ \\
\noalign{\smallskip}\hline
\end{tabular}
\end{table}

As it is shown in the Subsection \ref{ssapr0}, we obtain the equation (\ref{r0simple}).

\begin{equation}\label{r0simple} 
\mathcal{R}_0 = \frac{\mathcal{R}_0^h}{1-\mathcal{R}_0^r}
\end{equation}

In consequence,  $\mathcal{R}_0>1$ if and only if $\mathcal{R}_0^h + \mathcal{R}_0^r >1$. Using this equivalence we could determine whether $\mathcal{R}_0<1$ or $\mathcal{R}_0>1$ based on the values $\mathcal{R}_0^h, \mathcal{R}_0^r$.

\begin{figure}
\includegraphics[scale=0.5]{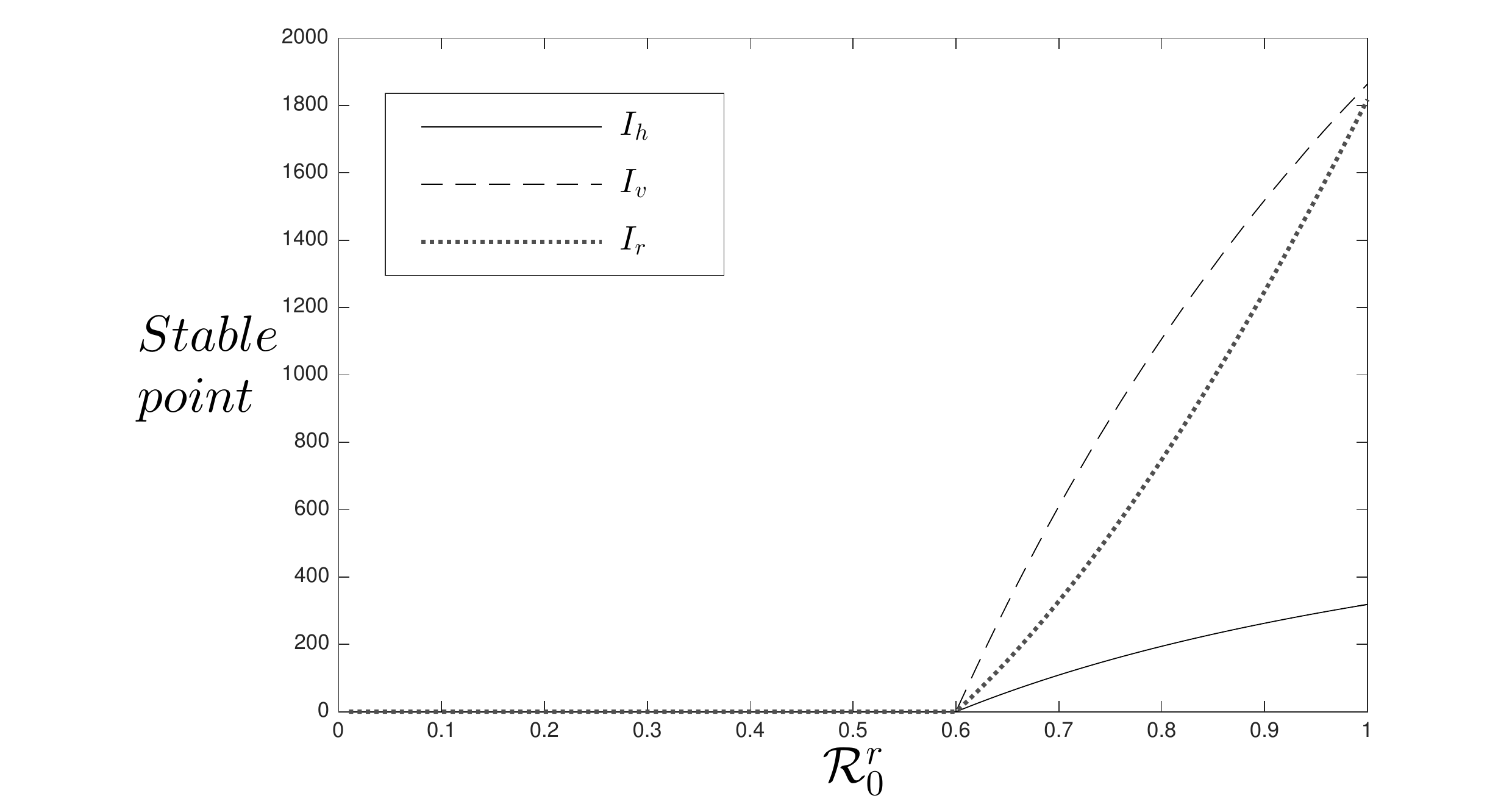}
\caption{Stable point of the system with one reservoir for $\mathcal{R}_0^h=0.4$. We consider values of the parameters for the reservoir $R_2$ taken from Table \ref{parameters}.}
\label{bif1}
\end{figure}

In Figure \ref{bif1}  we fix $\mathcal{R}_0^h=0.4$  and for different values of $\mathcal{R}_0^r$ we plot the stable points of the three infectious populations. In this figure we find a bifurcation in $1-\mathcal{R}_0^h = 0.6 $. In this example we observe how the weights of the cycles could be small but using  $\mathcal{R}_0^h, \mathcal{R}_0^r$ we can determine whether $\mathcal{R}_0<1$ or $\mathcal{R}_0>1$.

In the general scenario of the model presented in Section \ref{smodel}, we can obtain the same result in equation (\ref{r0simple}) using the values defined in Table \ref{tvalues2}.

\begin{table}
\caption{Umbral values defined from the systems formed by the state variables of the second column and the interpretations of the next generation matrix given by the third column}
\label{tvalues2}
\begin{tabular}{llll}
\hline\noalign{\smallskip}
Value & System form by & New infections \\ 
\noalign{\smallskip}\hline\noalign{\smallskip}
$\mathcal{R}_0$ & $H,V,R_i$, $i=1, \ldots, k$ & $H$ \\
$\mathcal{R}_0^r$ & $V, R$, $i=1, \ldots, k$ & $V$ \\
$\mathcal{R}_0^h$ & $H,V $& $H$ \\
$\mathcal{R}_0^{r_i}$ & $V, R_i$ & $V$ \\
\noalign{\smallskip}\hline
\end{tabular}
\end{table}

As it is shown in the Subsection \ref{ssapr0}, we obtain the equation (\ref{r0general}). If we also have that $\gamma_{ij} =0$, $i,j \in\{1,\ldots,k\}$, we obtain the equation (\ref{r0simple2}).

\begin{equation}
\mathcal{R}_0 = \frac{\mathcal{R}_0^h}{1-\mathcal{R}_0^r}
\label{r0general}
\end{equation}

\begin{equation}
\mathcal{R}_0^r  =  \sum_{j=1}^{k} \mathcal{R}_0^{r_j}
\label{r0simple2}
\end{equation}

In consequence,  $\mathcal{R}_0>1$ if and only if $\mathcal{R}_0^h + \mathcal{R}_0^r >1$. Furthermore, if  $\gamma_{ij} =0$, $i,j \in\{1,\ldots,k\}, i\neq j,$ we get that  $\mathcal{R}_0>1$ if and only if $\mathcal{R}_0^h +  \sum_{j=1}^{k} \mathcal{R}_0^{r_j} >1$. This equivalence lets us determine whether  $\mathcal{R}_0<1$ or $\mathcal{R}_0>1$ for small cycle weights, improving the informations obtained from the inequality in   \ref{umbralsimple2}.


\subsection{Application to Chagas disease} \label{ssrchagas}

\begin{figure}
\includegraphics[scale=0.5]{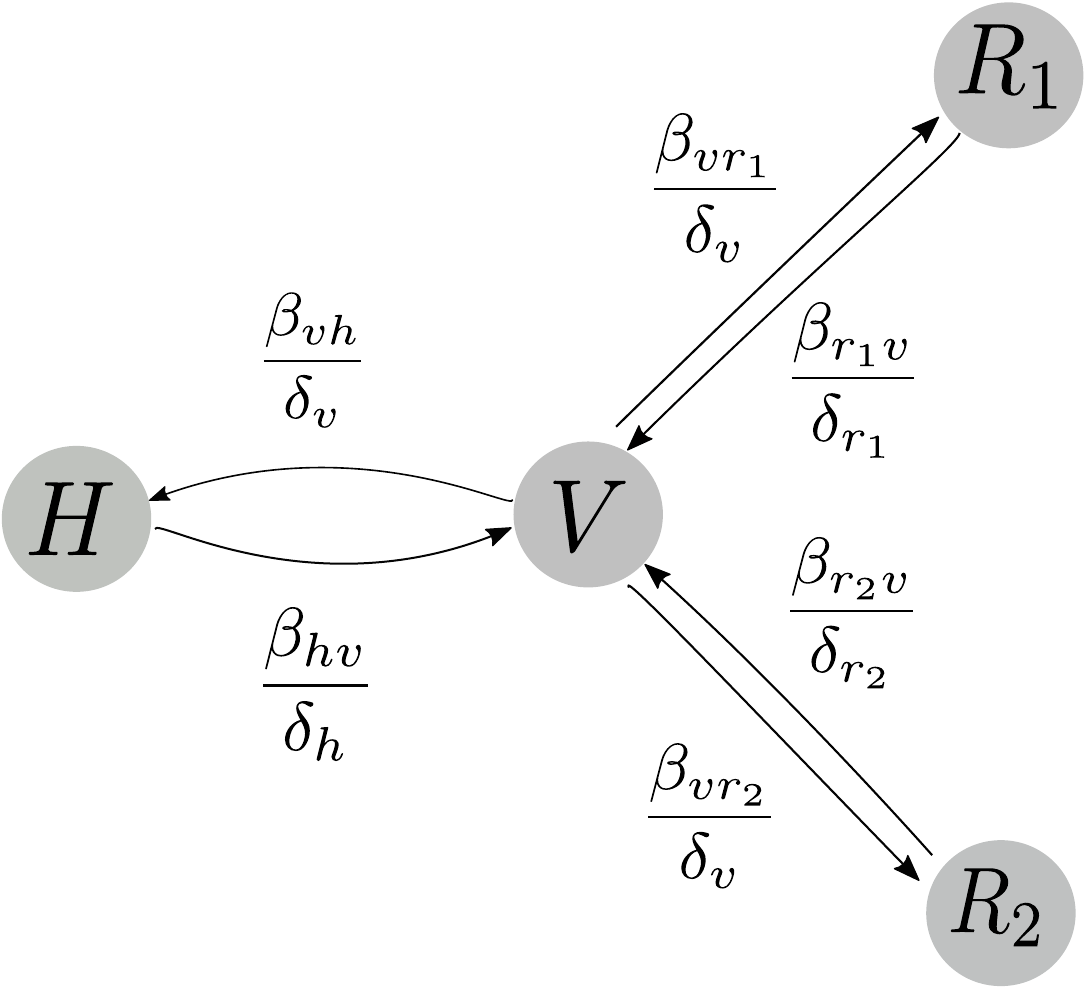}
\caption{Graph of Chagas model in \cite{cordovez2014environmental}. The species $R_1$ represents domiciliary reservoirs and the species $R_2$ represent the sylvatic reservoirs of the disease.}
\label{chagascycles}
\end{figure}

From \cite{cordovez2014environmental}, we can take some parameters for Chagas disease in Table \ref{parameters}. That paper considers a model with two kind of non-human host; the domiciliary hosts $R_1$ and the sylvatic hosts $R_2$. We consider the model presented in Section \ref{smodel} with two reservoirs  where there is no self infection ($ \nu_1 =0,\nu_2 =0$) and there is no transmission between reservoirs ($\gamma_{12}=0,\gamma_{21}=0$).  Figure \ref{chagascycles} is the graph of this model.

\begin{table}[H]
\caption{Parameters of the Chagas model in \cite{cordovez2014environmental}.}
\label{parameters}
\begin{tabular}{lllll} 
\hline\noalign{\smallskip}
Parameter              & Units                         & Estimate     \\ 
\noalign{\smallskip}\hline\noalign{\smallskip}
$\beta'_{hv}$ & Fraction of vectors infected by  & $1/(year*[H])$ & 0-1    \\ 
& one infectious human per year   & &\\
$\beta'_{vh}$ & Fraction of humans infected by    & $1/(year*[V])$ & $\beta'_{hv}$/100  \\
& one infectious vector per year   & &\\
$\beta'_{r_1v}$ & Fraction of vectors infected by  & $1/(year*[R_1])$ & 2 $\beta'_{hv}$ \\
& one infectious $R_1$ host per year   & &\\ 
$\beta'_{vr_1}$ & Fraction of $R_1$ hosts infected by  & $1/(year*[V])$ & $\beta'_{hv}$/10   \\
& one infectious vector per year   & &\\ 
$\beta'_{r_2v}$ & Fraction of vectors infected by & $1/(year*[R_2])$ & 1 $\beta'_{hv}$  \\ 
& one infectious $R_2$ host per year   & &\\ 
$\beta'_{vr_2}$ & Fraction of $R_2$ hosts infected by  & $1/(year*[V])$ & $\beta'_{hv}$/5   \\ 
& one infectious vector per year & &\\ 
$N_{r_1}$ & Number of $R_1$ individuals    & $[R_1]$ & 0.0005 $N_v$   \\
$N_{r_2}$ & Number of $R_2$ individuals     & $[R_2]$ & 0.001 $N_v$   \\
$N_{h}$ & Number of humans   & $[H]$ & 0.001 $N_v$         \\
$\delta_v $ & Mortality rate of vectors      & 1/year & 1         \\
$\delta_{r_1} $ & Mortality rate of $R_1$ hosts  & 1/year & 0.5         \\
$\delta_{r_2}$ & Mortality rate of $R_2$ hosts  & 1/year & 0.3         \\
$\delta_h$ & Mortality rate of humans  & 1/year & 0.015         \\
$\beta_{vh}$ & Number of human infections caused by& $[H]/(years*[V])$ & $N_h \beta'_{vh}$\\
& one infectious vector per unir of time & & \\
$\beta_{hv}$ & Number of vector infections caused by & $[V]/(years*[H])$ & $N_v \beta'_{hv}$\\
& one infectious human per unir of time &  & \\
$\beta_{vr_i}$ & Number of infections of reservoir $R_i$ caused by  & $[R_i]/(years*[V])$ & $N_{r_i} \beta'_{vr_i}$\\
& one infectious vector per unir of time & &\\
$\beta_{r_iv}$ & Number of vector infections caused by  & $[V]/(years*[R_i])$ & $N_v \beta'_{r_iv}$\\
& one infectious reservoir $R_i$ per unir of time & &\\
\noalign{\smallskip}\hline
\end{tabular}
\end{table}

If we define $\mathcal{R}_0^h,\mathcal{R}_0^{r_1},\mathcal{R}_0^{r_2}$ as in Table \ref{tvalues2}, from equations (\ref{r0general}) and (\ref{r0simple2}) we have that $\mathcal{R}_0 >1$ if and only if $\mathcal{R}_0^h + \mathcal{R}_0^{r_1} + \mathcal{R}_0^{r_2} >1$. Let us define $\mu'_{vh} := \frac{\beta'_{vh} \beta'_{hv}}{\delta_v \delta_h}$, $\mu'_{vr_1} := \frac{\beta'_{vr_1} \beta'_{r_1v}}{\delta_v \delta_{r_1}}$ and $\mu'_{vr_2} := \frac{\beta'_{vr_2} \beta'_{r_2v}}{\delta_v \delta_{r_2}}$. We have that $\mathcal{R}_0^h + \mathcal{R}_0^{r_1} + \mathcal{R}_0^{r_2} = N_v N_h \mu'_{vh} + N_v N_{r_1} \mu'_{vr_1} +  N_v N_{r_2} \mu'_{vr_2} $. Using the parameters in Table \ref{parameters} we obtain $\mu'_{vh} = \frac{2}{3}(\beta'_{hv})^2$, $\mu'_{vr_1} = \frac{2}{5}(\beta'_{hv})^2$ and $\mu'_{vr_2} = \frac{2}{3}(\beta'_{hv})^2$. In consequence,  $\mathcal{R}_0 >1$ if and only if 

\begin{equation}
N_{r_2} > - \frac{\mu'_{vr_1}}{\mu'_{vr_2}} N_{r_1} + (\frac{1}{N_v \mu'_{vr_2}} - \frac{N_h \mu'_{vh}}{\mu'_{vr_2}}) = -\frac{3}{5}N_{r_1} + (\frac{3}{2N_v \beta_{hv}^2} - N_h) 
\label{restreshold}
\end{equation}

\begin{figure}
\includegraphics[scale = 0.5]{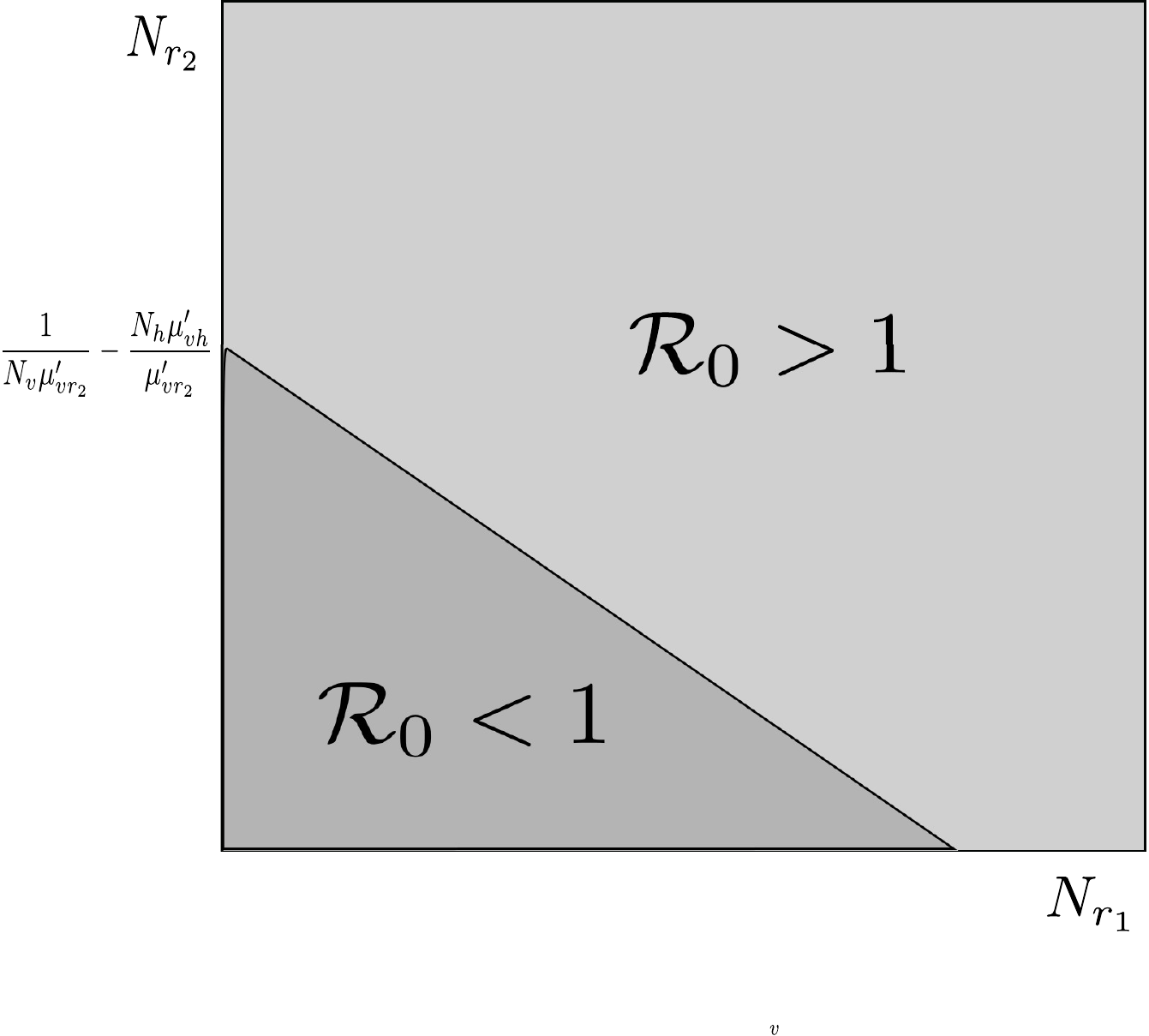}
\caption{Sign of $\mathcal{R}_0 -1$ depending on the number of reservoirs.}
\label{line}
\end{figure}

We must remark that $\mathcal{R}_0$ is always greater than $1$  if $\frac{1}{N_v \mu'_{vr_2}} - \frac{N_h \mu'_{vh}}{\mu'_{vr_2}}$ is negative. This is telling that if we would want to attack the disease, we first must control the vector. In the case that $\frac{1}{N_v \mu'_{vr_2}} - \frac{N_h \mu'_{vh}}{\mu'_{vr_2}} >0$, the number of reservoirs determine whether $\mathcal{R}_0>1$ according to the inequality in (\ref{restreshold}), as Figure \ref{line} shows.

\section{Conclusions} \label{sdiscusion}

Based on the model of Section \ref{smodel}, the endemicity of the disease in one reservoir could entail the endemicity of the disease in human population. We also conclude that human endemicity of a disease in our model could not be only explained considering the dynamics of the infection within an specific system of hosts. In an specific system, we could get a small basic reproductive number that does not explain the endemicity of the disease.  In our model, we observe how the basic reproductive numbers of the cycles between each  reservoir and the vector could be less than one separately. However, the sum of the effects of the reservoirs can lead to endemicity of the disease in all species. In consequence, we conclude that a large enough system of hosts that contribute to spread the infection must be identified to get rid of the endemicity of the disease. As an example of control of a disease, using  the data of Chagas disease we conclude that only dropping the abundancy of the reservoirs can not extinguish the disease. The abundancy of the vectors must be dropped under certain threshold for the intervention of the reservoirs to work.

\section{Appendix}\label{sappendix}

\subsubsection{Next generation matrix} \label{ssngm}

The basic reproductive number of an infectious disease $\mathcal{R}_0$  can be defined as the expected number of secondary cases produced in a susceptible population that are caused by an infectious individual. The NGM method lets us compute  $\mathcal{R}_0$  in an epidemiological model where the individuals are classified in different compartments and the dynamics of the size of those compartments is described by a system of ordinary differential equations (this method is explained in \cite{van2002reproduction}). The number that we get is a threshold for the local asymptotic stability of the disease free equilibrium $x_0$.

Let us assume that we have  $N=n+m$ types of individuals and that $\bar{x}= (x_1,\ldots, x_n)$ represents the number of individuals in infectious compartments $I_1, \ldots, I_n$ and $\bar{y}= (y_1,\ldots, y_m)$ represents the number of individuals in non-infectious compartments $J_1,\ldots, J_m$. In the model presented in Section \ref{smodel} we assume that all species populations are constant, so the system is only determined by the equations of the infectious compartments. For simplicity, we omit  the equations for non-infectious compartments in this  explanation. For a general exposition of the NGM, see  \cite{van2002reproduction}.  

Let us assume that the number of infectious individuals follows the system of equations in (\ref{xi}). In this system $f_i$ represents the new infections rate for the compartment $I_i$ and $v_i$ represents the rate of change of the size of this compartment due to other reasons, such as recovery, death or movement from other compartment due to causes different from new infection, like a disease stage. The functions $f_i$ and $v_i$ depend on the interpretation of which infectious individuals are regarded as new infections. Different interpretations will lead us to different versions of NGM. 

\begin{equation}
\frac{dx_i}{dt}= f_i(\bar{x},\bar{y})-v_i(\bar{x},\bar{y}), i=1,\ldots,n.
\label{xi}
\end{equation}

We define the disease free equilibrium (DFE)  as an equilibrium $x_0 = (\bar{x_0},\bar{y_0})$ of the system in (\ref{xi}) where $\bar{x} = (0, \ldots, 0)$. We define an endemic equilibrium as an equilibrium $x_0 = (\bar{x_0},\bar{y_0})$  where $\bar{x} \neq (0, \ldots, 0)$.

Let us assume that  $f_{ij}$ is the number of infections in the compartment $I_i$ that are caused by an individual in the compartment $I_j$ per unit of time in a susceptible population. In terms of the system in (\ref{xi}), we get $f_{ij} = \partial f_i(\bar{x},\bar{y})/\partial x_j |_{(\bar{x},\bar{y})=x_0}$. Let $F$ be the matrix $(f_{ij})_{n\times n}$. Let us also assume that $t_{jk}$ is the time that an individual from the compartment $I_k$ will be in the compartment $I_j$.  It turns out that $V^{-1}=(t_{ij})_{n\times n}$, where $V := (v_{ij})_{n \times n}$ and $v_{ij} = \partial v_i(\bar{x},\bar{y})/\partial x_j |_{(\bar{x},\bar{y})=x_0} $. If  $g_{ik}$ is the number of infections in the compartment  $I_i$ caused by an individual in the compartment in $I_k$ during its generation in a susceptible population, we should have $g_{ik} = f_{i1}t_{1k} + \ldots + f_{in}t_{nk} $. Each term $f_{ij}t_{jk} $ accounts for the infections caused by the individual that started in the compartment $I_k$ and spent a time $t_{jk} $ in the compartment $I_j$. Let $G$ be the matrix $(g_{ij})_{n\times n} = FV^{-1}$. We call $G$ the next generation matrix of the system in (\ref{xi}) with its respective interpretation contained in the functions $f_i,v_i, i=1,\ldots,n$. Finally, we define the basic reproductive number $\mathcal{R}_0$ (which depends on $G$) as the spectral radius of $G$, i.e.,   $\mathcal{R}_0=\rho(FV^{-1})$.

If the functions $f_i$ and $v_i$ satisfy the axioms $A1-A5$ presented in \\ \cite{van2002reproduction},  we obtain Theorem \ref{umbral} \\ (\cite[Theorem 2]{van2002reproduction}).

\begin{theorem}
Let $x_0$ be the DFE of (\ref{xi}). Then, $\mathcal{R}_0<1$ implies $x_0$ is locally asymptotically stable and $\mathcal{R}_0>1$ implies that $x_0$ es unstable.  
\label{umbral}
\end{theorem}

As an example, we consider the system of Figure \ref{feje1} and equations (\ref{eje1}) described in the Introduction. We obtain the basic reproductive numbers in Table \ref{texamplengm}.

\begin{table}
\caption{Basic reproductive numbers of the system of equations (\ref{eje1}). We define $\mu_h :=\frac{\nu_h}{\delta_h}$ and $\mu_{vh}:=\sqrt{ \frac{\beta_{hv}}{\delta_h}\frac{\beta_{vh}}{\delta_v}}$. }
\label{texamplengm}
\begin{tabular}{llllll}
\hline\noalign{\smallskip}
Interpretation  & $F$ & $V$ & $G = FV^{-1}$ & $\rho(FV^{-1})$   \\ 
(New infections) & & &  &    \\ 

\noalign{\smallskip}\hline\noalign{\smallskip}
$H$ & 
$\begin{pmatrix}
\nu_h &\beta_{vh} \\
0 &  0\\
\end{pmatrix}$ & 
$\begin{pmatrix}
\delta_h  & 0\\
-\beta_{hv} & \delta_v
\end{pmatrix}$ &
$\begin{pmatrix}
\mu_{h}+\mu_{vh}&\frac{\beta_{vh}}{\delta_v} \\
0&  0  \\
\end{pmatrix}$ &
$\mathcal{R}_0^2 = \mu_{h} + \mu_{vh}^2$ \\

$V$ & 
$\begin{pmatrix}
0 &0 \\
\beta_{hv} &  0\\
\end{pmatrix}$ & 
$\begin{pmatrix}
-\nu_h+\delta_h  & - \beta_{vh}\\
0 & \delta_v
\end{pmatrix}$ &
$\begin{pmatrix}
0 & 0 \\
\frac{\beta_{hv}}{\delta_h - \nu_h}&  \mu_{vh}^2/(1 -  \mu_{h})  \\
\end{pmatrix}$ &
$\mathcal{R}_0^3 = \mu_{vh}^2/(1 -  \mu_{h})$ \\

$H,V$ & 
$\begin{pmatrix}
\nu_h &\beta_{vh} \\
\beta_{hv} &  0\\
\end{pmatrix}$ & 
$\begin{pmatrix}
\delta_h  & 0\\
0 & \delta_v
\end{pmatrix}$ &
$\begin{pmatrix}
\mu_h &\frac{\beta_{vh}}{\delta_v} \\
\frac{\beta_{hv}}{\delta_h} &  0  \\
\end{pmatrix}$ &
 $\mathcal{R}_0^1 = \lambda^*$, where \\

&&&&$(\lambda^*)^2 - (\mu_h)\lambda^* - \mu_{vh}^2=0$.\\

\noalign{\smallskip}\hline
\end{tabular}
\end{table}

Using Theorem \ref{umbral} we have that  $ \mathcal{R}_0^1 >1 \Longleftrightarrow \mathcal{R}_0^2 >1 \Longleftrightarrow  \mathcal{R}_0^3 >1 \Longleftrightarrow $ $DFE$ is unstable. In consequence, in order to verify the possible endemicity of the disease we could consider interpretations that simplify calculations.


\subsection{Basic reproductive numbers of the model} \label{ssapr0}

For the general model defined in Section \ref{smodel} we consider the numbers  $\mathcal{R}_0$, $\mathcal{R}_0^{h}$, $\mathcal{R}_0^{r_i}$ and  $\mathcal{R}_0^{r}$ defined from the systems and interpretation in Table \ref{tvalues2}. We can get the equations (\ref{er0happ}) and (\ref{er0riapp}) in a similar way to the numbers presented in Table \ref{texamplengm} in the previous Subsection.

\begin{equation}\label{er0happ}
\mathcal{R}_0^h = \frac{\beta_{hv}\beta_{vh}}{\delta_h \delta_v}
\end{equation}

\begin{equation}\label{er0riapp}
\mathcal{R}_0^{r_i} = \frac{\beta_{r_iv}\beta_{vr_i}}{(\delta_{r_i} -\nu_i N_{r_i})\delta_v}
\end{equation}

As we only interpret infected humans as new infections, we get the NGM for $\mathcal{R}_0$ through the following matrices $F$ and $V$:

$$
F=
\begin{pmatrix}
0  &  \beta_{vh}             &  0    & 0  &  \dots                &  0 \\
0  &   0         &  0    & 0  &  \dots                &  0 \\
\vdots &                  &        & \ddots &      \ddots & \vdots \\
0  &   0         &  0    & 0  &  \dots                &  0 \\
\vdots &                  &        & \ddots &      \ddots & \vdots \\
0  &   0         &  0    & 0  &  \dots                &  0  \\  \end{pmatrix},
V =
\begin{pmatrix}
\delta_h    &     0         &  0    & 0  & \dots                &  0 \\
-\beta_{hv}  & \delta_v & -\beta_{r_1v}   &   -\beta_{r_2v}  &           \ldots              &  -\beta_{r_kv}    \\
0 &  -\beta_{vr_1}        & \delta_{r_1}-\nu_1    & -\gamma_{21}  &     \ldots     &-\gamma_{k1} \\
0 &  -\beta_{vr_2}     &  -\gamma_{12}  & \delta_{r_2}-\nu_2   &   \ddots     &-\gamma_{k2} \\
\vdots &                  &     \ddots &  \ddots    &    \ddots & \vdots \\
0 &  -\beta_{vr_k} &  -\gamma_{1k}  & -\gamma_{2k}    &  \ldots     &\delta_{r_k}-\nu_k \\  \end{pmatrix}
$$

Let us prove that $\rho(FV^{-1}) = \frac{\mathcal{R}_0^h}{1-\mathcal{R}_0^r}$. The adjugate matrix of $V$ enables us to obtain $V^{-1} = (t_{ij})_{(k+2)\times(k+2)}$. In particular, we are interested in the entry $t_{21}$ in (\ref{t21}), where $V_{i,j}$ denotes the matrix  that is obtained omitting the row $i$ and the column $j$ of $V$ and $K$ is the block matrix of $V$ formed by the entries $v_{ij}$ where $i=3,\ldots, k+2$ and $j=3,\ldots, k+2$.
 
\begin{equation}
t_{21} = \frac{-det(V_{1,2})}{det(V)}= \frac{\beta_{hv} det(K)}{\delta_h det(V_{1,1})}
\label{t21}
\end{equation}

Let $G= (g_{ij})_{(k+2)\times(k+2)}$ be $FV^{-1}$. Using (\ref{t21}), we get that:

\begin{equation}
\mathcal{R}_0=g_{11}=( \beta_{vh}  )(t_{21})= \frac{\beta_{hv} \beta_{vh} det(K)}{\delta_h det(V_{1,1})}.
\label{r0}
\end{equation}

On the other hand, when humans are not considered, we can obtain $\mathcal{R}_0^r$ from the spectral radius $\mathcal{R}_0^r = \rho(HW^{-1})$, where

$$H =
\begin{pmatrix}
0  &  \beta_{r_1v}   &   \beta_{r_2v} &     \ldots   &  \beta_{r_kv}   \\
0  &    0    & 0  &  \dots                &  0 \\
\vdots &                     & \ddots &     \ddots & \vdots \\
\vdots &                    & \ddots     &    \ddots & \vdots \\
0  &   0    & 0  & \dots                &  0  \\  \end{pmatrix}, W =
\begin{pmatrix}
 \delta_v &    0         &  0  &  \dots                &  0   \\
-\beta_{vr_1}  & \delta_{r_1}-\nu_1 & -\gamma_{21}  &     \ldots     &-\gamma_{k1}\\
 -\beta_{vr_2} &  -\gamma_{12} & \delta_{r_2}-\nu_2   &     \ldots     &-\gamma_{k2} \\
 \vdots &                  &     \ldots    &     \ddots & \vdots \\
-\beta_{vr_k}  &  -\gamma_{1k} & -\gamma_{2k} &  \ldots     & \delta_{r_k}-\nu_k \\  \end{pmatrix}.
$$

From the adjugate matrix of $W$, we have  that if  $W^{-1} = (w'_{ij})_{(k+1)\times(k+1)}$, then

\begin{equation}
 w'_{i1} = \frac{(-1)^{i+1}det(W_{1,i})}{det(W)}= \frac{(-1)^{i+1}det(W_{1,i})}{\delta_v det(K)}, i = 2, 3, \ldots, k+1
 \label{wi1}
\end{equation}

Let $L= (l_{ij})_{(k+1)\times(k+1)}$ be $HW^{-1}$. Using (\ref{wi1}), we get that:

\begin{equation}
\mathcal{R}_0^r=l_{11}=\sum_{j=1}^{k}( \beta_{r_jv})(w'_{(j+1)1})= \sum_{j=1}^{k}( \beta_{r_jv})(\frac{(-1)^{j}det(W_{1,j+1})}{\delta_v det(K)})
\label{r0r}
\end{equation}

Furthermore, we have that: 

\begin{equation}
det(V_{1,1}) =\delta_v det(K) + \sum_{j=1}^{k}  (- \beta_{r_jv})(-1)^{j+1}det(W_{1,j+1}) = \delta_v det(K)(1-\mathcal{R}_0^r ) 
\label{v11}
\end{equation}

Replacing (\ref{v11})  in (\ref{r0}), it turns out that: 

\begin{equation}
\mathcal{R}_0 = \frac{\beta_{hv} \beta_{vh} det(K)}{\delta_h det(V_{1,1})}= \frac{\beta_{hv} \beta_{vh}det(K)}{\delta_h \delta_v det(K)(1-\mathcal{R}_0^r )}= \frac{\mathcal{R}_0^h}{1-\mathcal{R}_0^r}
\label{finalr0}
\end{equation}

If we put $\gamma_{ij} =0$, we have $det(K) = \prod _i (\delta_{r_i}-\nu_i)$ and from equation (\ref{r0r}) we get that: 

\begin{equation}
\mathcal{R}_0^r = \sum_{j=1}^{k}( \beta_{r_jv} )(\frac{\beta_{vr_j} \prod_{i\neq j }(\delta_{r_i}-\nu_i )}{\delta_v det(K)})= \sum_{j=1}^{k} \frac{\beta_{r_jv} \beta_{vr_j}}{ \delta_v(\delta_{r_j}-\nu_j )} =  \sum_{j=1}^{k} \mathcal{R}_0^{r_j}
\label{nointeractionr0}
\end{equation}



\end{document}